\documentclass[prl,twocolumn,groupaddress,superscriptaddress,showpacs]{revtex4}

\usepackage{graphicx}
\usepackage{amssymb}
\usepackage[usenames]{color}

\begin{document}

\title{Evidence for a High-Temperature Disorder-Induced Mobility in Solid $^4$He }

\author{A. Eyal}
\affiliation{Department of Physics, Technion - Israel Institute of Technology, Haifa
32000, Israel}

\author{O. Pelleg}

\affiliation{Brookhaven National Laboratory, Condensed Matter and Materials Science, Upton, New York 11973-5000}

\author{L. Embon}
\affiliation{Department of Physics, Technion - Israel Institute of Technology, Haifa
32000, Israel}

\author{E. Polturak}

\affiliation{Department of Physics, Technion - Israel Institute of Technology, Haifa
32000, Israel}

\date{\today}

\begin{abstract}
We have carried out torsional oscillator experiments on solid $^4$He at temperatures between 1.3K and 1.9K. We discovered phenomena similar to those observed at temperatures below 0.2K, which currently are under debate regarding their interpretation in terms of supersolidity. These phenomena include a partial decoupling of the solid helium mass from the oscillator, a change of the dissipation, and a velocity dependence of the decoupled mass. These were all observed both in the bcc and hcp phases of solid $^4$He. The onset of this behaviour is coincidental with the creation of crystalline disorder but does not depend strongly on the crystalline symmetry or on the temperature.
\end{abstract}

\pacs {67.80 bd.}

\maketitle

Thirty odd years after being predicted \cite{Andreev1, Leggett}, and several years after some indirect experimental evidence \cite{Goodkind}, the discovery of Kim and Chan (KC) \cite{KC} of a possibly supersolid phase in hcp $^4$He led to an intense activity in order to better understand the nature of this new state. Several groups which repeated the KC experiment confirmed the basic effect, showing in addition that its magnitude depends strongly on the way the solid was prepared \cite{Reppy}. Recent experiments reveal additional interesting properties of this novel state. Among these are a possible heat capacity peak \cite{Chan} at the transition, an increase of the shear modulus \cite{Beamish} and of the relaxation time \cite{Davis} of the solid He when entering this state. In addition, some unusual mass flow was also observed in solid He below 1K \cite{Hallock}. Theoretical studies indicate that supersolidity is not expected in a perfect solid, but may be possible in a crystal containing structural defects \cite{Pollet, Ceperley}. Recently, it was suggested that two level systems can give rise to supersolidity \cite{Andreev2, Korshunov, Nussinov}. It is therefore quite clear that the crystalinity of the solid is an important issue. The question which initially interested us was whether effects similar to those observed in the hcp solid phase of $^4$He at low temperatures may be observed in the bcc solid phase as well. Bcc $^4$He exists only at temperatures approximately an order of magnitude higher than those where KC saw the effect.

Like other groups \cite{KC, Reppy, Davis, Kojima}, we use the torsional oscillator (TO) technique. Our method involves growing high quality single crystals of solid He. Single crystals of 1cm$^3$ in volume are grown inside an annular sample space of the TO. The inner radius of the annulus is 6.5 mm, the height is 10 mm, and its with is 2 mm. All internal corners of the sample space were intentionally rounded to enable solid to fill the cell without leaving residual pockets of liquid \cite{Balibar1,Balibar2}. He is admitted into the sample space through a heated filling line at the top of the TO structure. The oscillator is thermally connected to a $^3$He refrigerator through its torsion rod. Hence, the bottom of the cell inside the oscillator is the coldest internal surface and the solid nucleates there. Within our range (1.3K-1.9K), the resonant period of the oscillator itself does not depend on temperature. The loaded quality factor of the oscillator is around 500,000. Once the experimental cell is filled with solid He, we create structural disorder in the crystal. We then test how the appearance of this disorder affects the TO. We gained extensive experience in the growth and structural characterization of He crystals during our neutron scattering experiments \cite{Pelleg}.
The critical shear stress of solid He is very small. Hence, a single crystal of Helium becomes disordered (polycrystalline) if it is compressed or cooled while constrained by the rigid walls of the cell. Cooling generates stress due to the large thermal expansion of the crystal (10$^{-2}$ K$^{-1}$ at 1.6 K). We found that as long as there is any fluid in the cell, it acts as a buffer against thermal and mechanical shocks and the single crystal is stable. Only if the cell is completely filled with solid, the constrained crystal disorders under application of any stress such as described above.
To illustrate this process, in figure 1 we show neutron scattering experiment measurements of a bcc crystal. With some liquid in the neutron scattering cell, there is a single intense diffraction peak indicating that the solid is a single crystal. After the cell was filled with solid and the pressure step was applied, the diffraction pattern changed to that of a polycrystal composed of grains separated by low angle grain boundaries. The orientation of the polycrystal shifted spontaneously by a few degrees \cite{Pelleg2}.
Using the Laue technique \cite{Pelleg}, we found that in all cases where a crystal is disordered it becomes permeated by low angle grain boundaries, the density of which increases with the amount of stress applied to the crystal. These neutron scattering experiments were done using several different cells of different volume and shape, subjected to a wide variety of pressure and temperature changes. The kind of disorder generated in the solid He was always similar to what we show in figure 1. Hence, this result is quite general, and we take it to apply to our TO experiment as well.

\begin{figure}[]
\includegraphics[width=2 in]{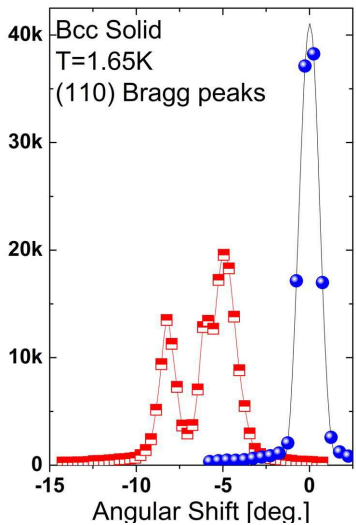} {}
\caption{Neutron diffraction scans of a bcc $^4$He crystal. A single crystal produces a sharp and intense diffraction peak (blue circles). The red squares show the effect of stress resulting in a breakup of the single crystal into slightly misoriented grains \cite{Pelleg2}. We note that this neutron scattering data was obtained in an independent experiment.}
\label{fig1}
\end{figure}

In our TO experiment, we simultaneously monitor the resonant period, P, the Q of the oscillator, and the pressure inside the cell. Pressure is measured using a capacitive pressure gauge attached to the top of the cell. Swelling of the sample cell with pressure increased the capacitance. The sensitivity of the gauge is 0.026pF/bar, with a mbar resolution, its response is linear and non-hysteretic over our working range, and its temperature dependance is negligible. Crystals were grown at a constant temperature and pressure on the melting curve by adding small amounts of He to the cell (Fig. 2). Over the duration of the growth, the temperature changed by less than ±2mK and the pressure at equilibrium by less than ±10mbar (out of 25-30bar). This procedure yields high quality crystals of uniform density.  The end of the filling process was marked by three indicators: the period, P, stops changing, Q increases rapidly, and the pressure inside the cell increases.  This initial increase of the pressure also serves to solidify the liquid inside the filling line, decoupling the cell from any subsequent variations of the external pressure.  The measured pressure and temperature of the cell at equilibrium serve to locate the coordinates of the solid on the phase diagram.

We now discuss the response of the oscillator to stress applied to the crystal.  The data shown below are representative of experiments done with more than 30 crystals grown at different temperatures, and with two different TOs.
First, if the growth was stopped with some residual liquid in the cell, the period, P, stopped changing and remained constant over days. This indicates that the crystals as grown are of high quality and no annealing takes place. With a cell full of solid, the behaviour changes drastically. In Figure 2, we show the period, P, at the final stage of crystal growth and after the introduction of disorder. As the crystal grows, P and Q increase, until the cell is completely full. As the crystal becomes disordered due to additional compression, P decreases almost immediately, indicating that mass is decoupling from the oscillator. In the following, we use the term "mass decoupling" in conjunction with the period change, although it more probably results from the solid at grain boundaries and interfaces becoming mobile. Simultaneously with the onset of the period change, the Q of the oscillator strongly decreases (up to a factor of 30). Under steady conditions, this process is completed within 10 min up to 2 hours, depending on temperature. After that, P and the Q of the oscillator containing a disordered crystal stay constant for as long as the ambient conditions remain steady.  The change in the moment of inertia of any given crystal due to the introduction of disorder was 5\%-30\% of the total moment of inertia attributed to solid $^4$He. Here $\Delta I/I_{solid}\approx(P_{decoupled}-P_{full})/(P_{empty}-P_{full})$ . The magnitude of this change is not intrinsic, as it seems to depend on the details of the procedure by which we disorder the crystal.  We found that the very appearance of mass decoupling does not depend on the crystalline symmetry. Within the temperature range of our experiment, between 1.3K and 1.9K, we grew bcc crystals, hcp crystals grown from the superfluid (below 1.464K), and hcp crystals grown from the normal fluid (above 1.772K).  All of these showed mass decoupling. The only systematic difference between these was that it took about 10 times longer to grow a single crystal from the normal fluid than from the superfluid, presumably because of the longer time it takes the latent heat to leave the system.

\begin{figure}[]
\includegraphics[width=3.375 in]{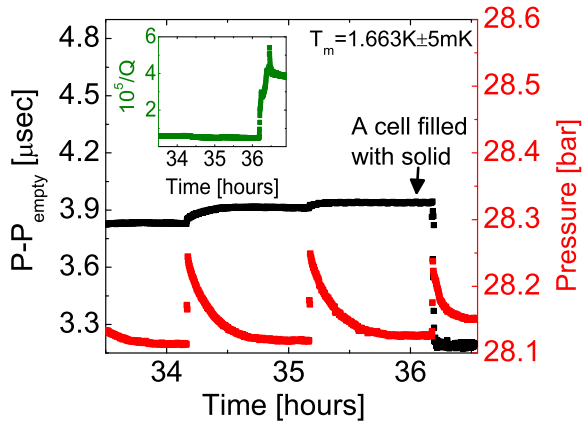} {}
\caption{Final stages of crystal growth. Time is measured from the beginning of the growth. Small quantities of He are periodically added, momentarily increasing the pressure inside the cell (red symbols) by  130 mbar over the melting pressure (28.13$\pm$0.01 bar). As He solidifies in the cell the overpressure decays to zero and the period, P, of the oscillator increases (black symbols). A small pressure step applied to a filled cell at t=36.2 hours compresses the solid enough to disorder the crystal. As a result, P decreases, equivalent to a mass decoupling of 19\%. The inset shows Q$^{-1}$ of the oscillator, which increases upon disordering the crystal. P of an empty cell is 2936.76$\mu$sec}
\label{fig2}
\end{figure}

Figure 3 shows the temperature dependence of the mass decoupling for an hcp crystal grown from the normal fluid. Cooling the disordered crystal leads to a further increase of the amount of decoupled mass. In parallel, the Q of the oscillator increases, signifying a decrease in dissipation. In Figure 4 we show a similar experiment for a bcc crystal grown at 1.64K. Also here we observe mass decoupling following the disordering of the crystal. During subsequent cooldown, the pressure inside the cell follows the bcc-hcp transition line.  Once the cell reaches the triple point at T=1.464K, P=26.36 bar, liquid appears in the cell and the decoupling effect vanishes. As the cell is cooled further along the melting curve, the liquid re-solidifies and the crystal disorders again. The decoupled mass fraction at lower temperatures is smaller, which may be due to the lower thermal expansion at these temperatures (this would cause a smaller stress which breaks the crystal less effectively). With each crystal, our initial set of data is taken during cooling from the temperature at which it was grown. We found no essential difference between data taken during cooling or warming, in contrast to what was observed at some recent low temperature experiments \cite{Davis}.

\begin{figure}[]
\includegraphics[width=3.375 in]{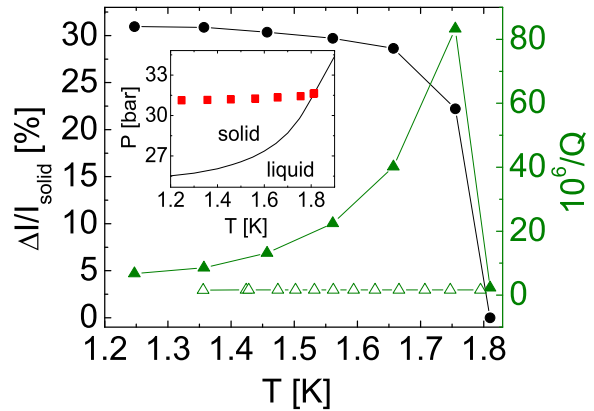} {}
\caption{Temperature dependence of the decoupled mass fraction (black circles) and of the dissipation Q$^{-1}$ (green triangles) for an hcp crystal grown from the normal fluid at 1.81K. Initially, as the crystal is disordered, about 21\% of the mass decouples and the Q of oscillations decreases. Upon cooling, both the Q of oscillation and the amount of decoupled mass increase. The inset shows the pressure inside the cell. The solid line in the inset is the melting curve \cite{Grilly}. The open triangles show the temperature dependence of the dissipation of an empty cell.}
\label{fig3}
\end{figure}

\begin{figure}[]
\includegraphics[width=2.5 in]{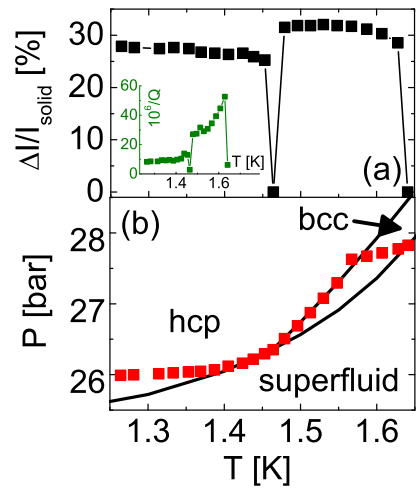} {}
\caption{(a) Temperature dependence of the decoupled mass of a He crystal grown and disordered at 1.64K. Initially the crystal is bcc. During cooldown the crystal gradually converts to an hcp one. The inset shows the dissipation (Q$^{-1}$) of the oscillator while cooling. (b) Temperature dependence of the pressure inside the cell during the cooldown (red symbols). The solid line shows the phase diagram  \cite{Grilly}. Note that once the crystal reaches the lower triple point (1.464K), liquid appears in the cell, the decoupled mass fraction vanishes and Q$^{-1}$ decreases. }
\label{fig4}
\end{figure}

We took particular care to ensure that the behaviour observed is not due to the presence of some residual fluid inside the cell. Our in-situ pressure data shown in Figs. 3 and 4 clearly indicate that the He inside the cell is in the part of the phase diagram where only solid is stable.  A second indication is that mass decoupling was never observed when there was any fluid present in the cell. As seen in Fig. 4, the moment any fluid appears in the cell, the mass decoupling effect disappears, the crystal anneals, and the moment of inertia and Q return to the values of a cell filled with a high quality single crystal. The wide temperature range over which mass decoupling was observed further indicates that the phenomena is unrelated to the nature of the fluid phase from which the crystals were grown. We stress the importance of designing the TO to have no sharp corners. Optical studies of He crystals \cite{Markovitz, Balibar1} show that due to the non wetting nature of the solid \cite{Balibar2} a cell having sharp corners is practically impossible to fill with solid without leaving some trapped liquid. With any liquid inside the cell, the effects reported here are not observable. In experiments done with the same cell, but with sharp corners, we did not see any of the effects described here.

In addition, we have also performed measurements of the mass decoupling fraction as a function of the tangential rim velocity of the oscillator. Figure 5 shows the results of these measurements. Changes of P of a cell filled with fluid, or containing a single (ordered) crystal, are essentially identical to changes measured with an empty cell. Hence, single crystals behave as a perfect rigid body. In contrast, P of a cell containing a disordered crystal increases by a significant amount with the velocity, i.e. the decoupled mass becomes smaller. The data obtained while increasing or decreasing the velocities are the same, unlike \cite{Kojima}. Except for the velocities being an order of magnitude higher, our results resemble those by other groups obtained at low temperatures.

\begin{figure}[]
\includegraphics[width=3.375 in]{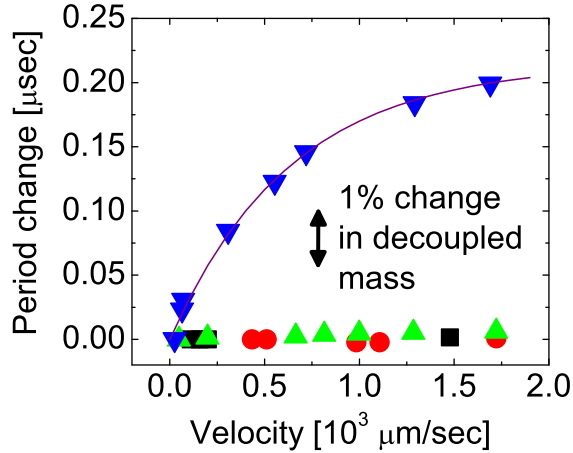} {}
\caption{Velocity dependence of the mass decoupling. For convenience, all the data sets were shifted to have a common origin. Blue triangles: disordered bcc crystal with 13.6\% decoupled mass fraction (at small velocities). For comparison, we show data for: empty cell (black squares); a cell filled with liquid (red circles); and a cell filled with a single crystal (green triangles). The solid line is a fit to an exponential function.}
\label{fig5}
\end{figure}

In summary, in our high temperature TO experiment, we observe behaviour which is qualitatively very similar to that seen at low temperatures \cite{KC,Kojima,Davis,Reppy}. The appearance of the effects seen in our experiment is directly correlated with the creation of disorder (grain boundaries) in a single crystal. The mass decoupling seems to be insensitive to the growth temperature between 1.3K and 1.9K. Although our data clearly show no bulk liquid inside the cell, nevertheless some form of mobility can be associated with the grain boundaries. We remark that if the grain boundaries of the crystal are mobile, then it would be possible to observe a very large mass decoupling, even though the volume of the grain boundaries is small. The solid at grain boundaries is naturally disordered. Hence, the effect should not depend strongly on the crystalline symmetry, which is consistent with our observations. Although the phenomena we observed are qualitatively similar to those seen at low temperatures, this does not imply that there is necessarily a common origin. Therefore, theories developed in conjunction with crystalline defects for the low temperature experiments \cite{Pollet,Andreev2, Korshunov, Nussinov} may or may not be applicable to our results.

We thank S. Hoida, A. Post, and L. Yumin for their help with the experiment. This work was supported by the Israel Science Foundation and by the Technion Fund for Research. \\
Author Information: satanan@tx.technion.ac.il

\end{document}